\documentclass[11pt]{article}
\pdfoutput=1

\usepackage{jheppub} 
\usepackage{slashed}
\usepackage[T1]{fontenc}
\usepackage{amsmath}
\usepackage{amssymb}
\usepackage{esint}
\usepackage{lmodern}
\usepackage{slashed}
\usepackage{dsdshorthand}
\usepackage{tikz} 
\usepackage[bottom]{footmisc}
\usepackage{mathdots}
\usetikzlibrary{arrows.meta} 
\usepackage{tabularx}
\usetikzlibrary{arrows}
\usepackage{subcaption}
\usepackage{float}

\def\bz{\bar{z}}

\newtheorem*{theorem*}{Theorem}



\title{Critical Slowing Down and the Crossover to Quantum Scrambling in Holographic CFTs}

\author[a]{Mozib Bin Awal}
\author[a,b]{and Prabwal Phukon}

\affiliation[a]{Department of Physics, Dibrugarh University - Assam, 786004, India}
\affiliation[b]{Theoretical Physics Division, Centre for Atmospheric Studies, Dibrugarh University, Assam}

\emailAdd{rs\_mozibbinawal@dibru.ac.in}
\emailAdd{prabwal@dibru.ac.in}

\abstract{We investigate critical relaxation dynamics and their connection to quantum information scrambling in the conformal field theory dual to a Reissner-Nordstr\"om anti-de Sitter (RN-AdS) black hole. Employing the holographic dictionary, we formulate the thermodynamics in terms of boundary CFT variables and study the stochastic evolution of the reduced horizon radius using Langevin and Fokker-Planck formalisms. We demonstrate pronounced critical slowing down in the dual CFT, characterized by an enhancement of the autocorrelation time and stochastic fluctuations, accompanied by a suppression of the lowest nonzero Fokker-Planck eigenvalue. The autocorrelation time exhibits a universal power-law divergence, $\tau\sim|\epsilon|^{-\Delta}$, with $\Delta\simeq 2/3$, consistent with the analytical prediction and previous bulk gravitational results.  Going beyond the conventional single-parameter analysis, we incorporate variations of the CFT central charge and identify a combined scaling variable involving temperature and central charge, demonstrating that the leading relaxation dynamics retain the same critical exponent along different trajectories in the two-dimensional parameter space. Finally, we explore the interplay between thermodynamic relaxation and quantum information scrambling by comparing the relaxation timescale with the fast-scrambling timescale. The comparison reveals a crossover between thermodynamic relaxation and quantum scrambling for which we derive the corresponding crossover scale analytically and find that, in the large-central-charge limit, it behaves as $\hat{\epsilon}\sim(C\ln C)^{-3/2}$. We also verify this numerically using the full nonlinear free-energy landscape. Our results connect critical dynamical scaling, finite-central-charge effects, and quantum scrambling within the holographic CFT framework.}

\begin{document} 

\maketitle
\section{Introduction}\label{sec1}
Black hole thermodynamics has emerged as one of the most profound and central topics in modern gravitational physics. The pioneering works of Bekenstein, Hawking, Bardeen, and others established a thermodynamic framework for black holes, within which fundamental thermodynamic quantities such as entropy and temperature acquire well-defined physical meanings \cite{Phys,bekens,Hawking,Hawking2,Bardeen}. These developments naturally raised the intriguing question of whether black holes can exhibit genuine thermodynamic phenomena, such as phase transitions and critical behaviour, analogous to those encountered in conventional thermodynamic systems. In this context, black hole phase transitions were first investigated by P.~C.~W.~Davies \cite{Davies} and P.~Hut \cite{Hut}. 

The study of Anti-de Sitter (AdS) black holes acquired renewed significance following Maldacena's formulation of the AdS/CFT correspondence in 1997 \cite{Maldacena}. A further conceptual development came with the introduction of the ``extended phase space'' formalism, in which the cosmological constant $\Lambda$ is promoted to a thermodynamic pressure. This framework revealed a remarkably rich phase structure for black holes, including phase behavior analogous to that of conventional thermodynamic systems, most notably the van der Waals fluid \cite{Kubiz,Hawkpage,Cai,Kastor,Dolan,Dolan2,Dolan3,Kubizna,Xu,Xu2,Zhang}. Traditionally, black hole phase transitions have been investigated primarily through the analysis of thermodynamic free energies. In recent years, however, several alternative approaches have been developed to probe the phase structure of black holes. Notable examples include thermodynamic geometry \cite{Ruppeiner:2012uc,Miao:2017cyt,Guo:2019oad}, thermodynamic topology \cite{Wu:2022whe,Liu:2022aqt,Fan:2022bsq}, and, more recently, approaches based on Lyapunov exponents \cite{first,le,Awal}.

The concept of a free energy landscape provides a powerful framework for investigating phase transitions across a wide range of physical, chemical, and biological systems, as it offers a means of connecting macroscopic phase behaviour with the underlying dynamical processes \cite{protein,bio,Golden,noneq}. In particular, it has been proposed that the kinetics of phase transitions can be understood by studying the stochastic dynamics of the relevant order parameter in the presence of thermal fluctuations. Such dynamics can be formulated in terms of a Fokker - Planck equation defined over the corresponding free energy landscape \cite{p1,p2,p3,p4}. Despite its successful application to various conventional systems, this framework has only recently been employed to investigate the dynamical aspects of black hole phase transitions. An important development in this direction was presented by Li and Wang \cite{Li}, who investigated black hole phase transitions from the perspective of the free energy landscape. Within this framework, they examined the Hawking--Page transition in both Einstein and massive gravity. The approach was subsequently applied to the small--large black hole phase transition in Reissner Nordstr\"om - Anti-de Sitter (RN-AdS) black holes \cite{Li2}, whose phase structure exhibits a close correspondence with the liquid-gas transition of the van der Waals system. Numerical analyses of the corresponding Fokker-Planck equation demonstrated that stochastic transitions between the small and large black hole phases can proceed in both the forward and reverse directions. The free energy landscape formalism has since been generalized to a variety of other black hole configurations \cite{fp1,fp2,fp3,fp4,fp5,fp6}. In this framework, the order parameter is regarded as a stochastic dynamical variable subject to thermal fluctuations. Its evolution can equivalently be described by a Langevin equation containing dissipative, deterministic driving, and stochastic noise contributions, or by the corresponding Fokker-Planck equation governing the probability distribution of the system over the free energy landscape. This perspective provides a dynamical description of black hole phase transitions and establishes a direct connection between the structure of the free energy landscape and the underlying transition kinetics. In particular, these studies have highlighted the crucial roles played by the temperature and the height of the free energy barrier in controlling the transition dynamics.

More recently, considerable attention has been devoted to the dynamical description of black hole phase transitions. In particular, Ref.~\cite{slowing} demonstrated that the approach to the critical and spinodal points is accompanied by a pronounced critical slowing down, manifested through a substantial growth of both the autocorrelation time and the fluctuations of the order parameter. In their treatment, the order parameter was regarded as a stochastic degree of freedom evolving within the free energy landscape, with its dynamics described either by a Langevin equation or, equivalently, by the corresponding Fokker-Planck equation \cite{slowing}. The observed slowing down was attributed to the flattening of the free energy landscape near the critical regime and was further substantiated by examining the smallest eigenvalue of the associated Fokker-Planck operator. 

So far, a comparable critical slowing down analysis, including a study of the universal scaling law, has not been carried out in the dual CFT. This is of interest because the CFT description introduces a new control parameter, the central charge. We consider a charged AdS black hole and map the bulk thermodynamic quantities to boundary CFT variables using the holographic dictionary, study the slowing down near the critical point, and test the universality of the dynamical exponent. Additionally, varying the temperature and the central charge together, we show that the relaxation time is governed by a single scaling variable, a linear combination of the reduced temperature and the deviation of the central charge from its critical value. We also compare the relaxation time with the Sekino-Susskind scrambling time \cite{Maldacena:2015waa,Shenker}, argue that the two cross at a particular reduced temperature, derive an analytical expression for this crossover, and verify it numerically.

The paper is organized as follows. Section~\ref{section2} provides a brief review of holographic thermodynamics. In Section~\ref{section3}, we investigate critical slowing down in the dual CFT. Section~\ref{section4} examines the universal dynamical scaling behaviour, including the role of the central charge as an additional control parameter. In Section~\ref{section5}, we study the crossover between the relaxation time and the scrambling time. Finally, Section~\ref{section6} presents our conclusions and discusses possible directions for future work.

\section{Holographic Thermodynamics: A Review}\label{section2}

The thermodynamic description of asymptotically Anti-de Sitter (AdS) black holes can be generalized by treating the cosmological constant as a thermodynamic variable. In this extended phase-space formulation, the cosmological constant is associated with the pressure of the gravitational system,
\begin{equation}
P=-\frac{\Lambda}{8\pi G_N},
\qquad
\Lambda=-\frac{d(d-1)}{2L^2}
\label{eq2.1}
\end{equation}
where $L$ denotes the AdS curvature radius and $d$ is the spacetime dimension. This reinterpretation of $\Lambda$ leads to an enlarged thermodynamic phase space and allows black holes to exhibit a rich variety of phase phenomena analogous to those of ordinary thermodynamic systems.

A further extension can be obtained by allowing Newton's constant $G_N$ to vary. From the viewpoint of gravitational theory, $G_N$ plays the role of a coupling constant and its variation provides a natural extension of the thermodynamic phase space. It also allows one to organize the corresponding generalized first law and Smarr relation in a unified manner. For Einstein--Maxwell theory with a negative cosmological constant, the generalized first law can be expressed as
\begin{equation}
dM=\frac{\kappa}{8\pi G_N}dA+\Phi\,dQ+\frac{\Theta}{8\pi G_N}d\Lambda-(M-\Phi Q)\frac{dG_N}{G_N}
\label{eq2.2}
\end{equation}
while the associated Smarr relation takes the form
\begin{equation}
M=\frac{d-1}{d-2}\frac{\kappa A}{8\pi G_N}+\Phi Q-\frac{1}{d-2}\frac{\Theta\Lambda}{4\pi G_N}
\label{eq2.3}
\end{equation}
Here, $M$ represents the black hole mass, $Q$ is the electric charge, $\Phi$ is its conjugate electric potential, and $\kappa$ denotes the surface gravity. The quantity $\Theta$ is thermodynamically conjugate to the cosmological constant. It admits a geometric interpretation that can be formulated in terms of an appropriately defined volume associated with the gravitational configuration.

Upon identifying the cosmological constant with the thermodynamic pressure through Eq.~\ref{eq2.1}, the contribution involving the variation of $\Lambda$ can be rewritten in the familiar $VdP$ form. In particular, at fixed $G_N$, one may define the thermodynamic volume as
\begin{equation}
V=-\Theta
\label{eq2.4}
\end{equation}
such that the generalized first law reduces to
\begin{equation}
dM=TdS+\Phi\,dQ+VdP
\label{eq2.5}
\end{equation}
where we have used the standard relation
\begin{equation}
T=\frac{\kappa}{2\pi}
\label{eq2.6}
\end{equation}
The corresponding Smarr relation becomes
\begin{equation}
M=\frac{d-1}{d-2}TS+\Phi Q -\frac{2}{d-2}PV
\label{eq2.7}
\end{equation}

The extended phase-space description is particularly useful for studying the phase structure of AdS black holes. In this framework, the mass $M$ is naturally interpreted as the enthalpy of the black hole rather than its internal energy, with the thermodynamic volume $V$ playing the role conjugate to the pressure $P$. This perspective provides the basis for investigating black hole phase transitions, critical phenomena, and their possible correspondence with phase transitions in the dual conformal field theory.

The extended thermodynamic description also raises an important issue concerning the interpretation of the thermodynamic variables. In particular, when Newton's constant $G_N$ is allowed to vary, the contributions involving $\Lambda$ and $G_N$ in the first law cannot be expressed solely through a single variation of the ratio $\Lambda/G_N$. Equation~\ref{eq2.2} can instead be recast in the form
\begin{equation}
dM
=
\frac{\kappa}{8\pi}d\left(\frac{A}{G_N}\right)
+\Phi\,dQ
+\frac{\Theta}{8\pi}d\left(\frac{\Lambda}{G_N}\right)
-
\left(
M-\frac{\kappa A}{8\pi G_N}
-\Phi Q
-\frac{\Theta\Lambda}{8\pi G_N}
\right)
\frac{dG_N}{G_N}
\label{eq2.8}
\end{equation}
This expression makes explicit that varying $G_N$ introduces an additional thermodynamic contribution that is not absorbed into the conventional $(P,V)$ description.

Moreover, the interpretation of the black hole mass as enthalpy, which is natural within the extended phase-space approach, differs from the usual identification of mass with the internal energy. While the $(P,V)$ formulation provides a useful description of black hole phase transitions, its applicability to holographic thermodynamics is more subtle when $G_N$ is treated as a variable. In particular, the additional $dG_N$ contribution in Eq.~\ref{eq2.8} is not naturally encoded within the standard pressure--volume formulation. This observation motivates considering alternative thermodynamic variables that are more directly connected to the holographic description and, in particular, to the parameters characterizing the dual conformal field theory \cite{cft1,cft2,cft3,cft4,cft5,cft6}. Recent developments have further explored this issue and proposed extensions of the holographic thermodynamic framework \cite{cft7}.

A more natural formulation from the holographic perspective is obtained by replacing the bulk pressure with the central charge of the dual conformal field theory. In the AdS/CFT correspondence, variations of the bulk cosmological constant can be related to variations of the number of degrees of freedom of the boundary theory, conventionally characterized by the central charge $C$ (or, equivalently, by $N$) \cite{cft1,cft2,cft4}. For Einstein gravity, the central charge scales as
\begin{equation}
C \sim \frac{L^{d-1}}{G_N}
\end{equation}
where $L$ is the AdS curvature radius. Consequently, varying $C$ corresponds to correlated variations of the bulk parameters $L$ and $G_N$. The quantity conjugate to $C$ can therefore be interpreted as a chemical potential $\mu$, sometimes referred to as the ``color susceptibility'' \cite{cft2}. This provides a thermodynamic description in which the number of degrees of freedom of the dual field theory becomes a genuine thermodynamic variable.

The correspondence between the bulk and boundary thermodynamic descriptions can be made explicit by rewriting the bulk first law in terms of the holographic variables. Using
\begin{equation}
\frac{d\Lambda}{\Lambda}=-2\frac{dL}{L}
\end{equation}
and eliminating $\Theta$ through the corresponding Smarr relation, the bulk first law can be expressed as
\begin{equation}
dM=\frac{\kappa}{2\pi}
d\left(\frac{A}{4G_N}\right)
+\frac{\Phi}{L}d(QL)
-\frac{M}{d-1}\frac{dL^{d-1}}{L^{d-1}}+\left(M-\frac{\kappa A}{8\pi G_N}-\Phi Q\right)\frac{d(L^{d-1}/G_N)}{L^{d-1}/G_N}
\label{eq2.11}
\end{equation}

The holographic dictionary then relates the bulk thermodynamic quantities to those of the boundary CFT according to
\begin{equation}
E=M,
\qquad
\widetilde{\Phi}=\frac{\Phi}{L},
\qquad
\widetilde{Q}=QL,
\qquad
\mathcal{V}\sim L^{d-1},
\qquad
C\sim\frac{L^{d-1}}{G_N}
\label{eq2.12}
\end{equation}
Here, $E$, $\widetilde{\Phi}$, $\widetilde{Q}$, $\mathcal{V}$, and $C$ denote the energy, chemical potential associated with the conserved charge, conserved charge, spatial volume, and central charge of the dual CFT, respectively. With the standard identifications of the temperature and entropy, Eq.~\ref{eq2.11} takes the form of an extended first law for the boundary theory,
\begin{equation}
dE=T\,dS+\widetilde{\Phi}\,d\widetilde{Q}-p\,d\mathcal{V}+\mu\,dC
\label{eq2.13}
\end{equation}
where the chemical potential associated with the central charge is given by
\begin{equation}
\mu=\frac{1}{C}\left(E-TS-\widetilde{\Phi}\widetilde{Q}\right),
\qquad
p=\frac{1}{d-1}\frac{E}{\mathcal{V}}
\label{eq2.14}
\end{equation}

Thus, the bulk variation of $G_N$ and $L$ is encoded on the boundary through a variation of the central charge $C$. In this formulation, the thermodynamic degrees of freedom associated with the number of colors of the dual field theory are incorporated directly into the first law. The relation for $\mu$ can also be understood as the Euler relation for the deconfined large-$N$ CFT and is holographically related to the generalized Smarr relation \cite{cft2,Visser}. The boundary pressure $p$ is consistent with the equation of state of a conformal field theory in $d$ dimensions.

More generally, the spatial volume of the boundary theory need not be tied to the AdS curvature scale. Introducing an independent boundary curvature radius $\mathcal{R}$ allows the CFT volume and the central charge to be varied independently, with $\mathcal{V}$ controlled by $\mathcal{R}$ while $C$ continues to depend on the bulk combination $L^{d-1}/G_N$. The holographic matching between the bulk and boundary first laws is preserved in this more general setting \cite{Visser}.

The relevant references are the following \cite{cft1,cft2,cft3,cft4,cft5,cft6,cft7}. Then subsequently, Visser proposed that holographic thermodynamics requires a chemical potential for colour in \cite{Visser}.

\subsection{Extended thermodynamics of charged AdS black holes}
We revisit the spherically symmetric Reissner-Nordstrom AdS black hole in $d+1$ dimensions with a negative cosmological constant given by the action
\begin{equation}
I=\frac{1}{16\pi G_N}\int d^{d+1}x\sqrt{-g}\left(R-2\Lambda-F^2\right)
\end{equation}
with $F=dA$ being the electromagnetic field strength. The metric in static coordinates is given by
\begin{equation}
ds^2=-f(r)dt^2+\frac{dr^2}{f(r)}+r^2d\Omega^2_{d-1}
\end{equation}
With the lapse function 
\begin{equation}
f(r)=1+\frac{r^2}{L^2}-\frac{m}{r^{d-2}}+\frac{q^2}{r^{2d-4}}
\end{equation}
The parameters appearing in the black hole solution can be expressed in terms of the conserved ADM mass $M$ and electric charge $Q$ as
\begin{equation}
M=\frac{(d-1)\Omega_{d-1}}{16\pi G_N}\,m
\label{eq2.18}
\end{equation}
and
\begin{equation}
Q=\frac{(d-1)\Omega_{d-1}}{8\pi G_N}\,\alpha q,\qquad\alpha=\sqrt{\frac{2(d-2)}{d-1}}
\label{eq2.19}
\end{equation}
where $\Omega_{d-1}$ denotes the volume of the unit $(d-1)$-dimensional sphere, while $m$ and $q$ are the mass and charge parameters characterizing the solution. Choosing the gauge such that the temporal component of the gauge potential vanishes at the event horizon, $A_t(r_h)=0$, the corresponding electrostatic potential is given by
\begin{equation}
\Phi=\frac{1}{\alpha}\frac{q}{r_h^{d-2}}
\label{eq2.20}
\end{equation}
where $r_h$ denotes the event horizon radius. It is determined by the largest positive root of the horizon equation $f(r_h)=0$. Consequently, the mass parameter can be written in terms of the horizon radius and the charge parameter as
\begin{equation}
m=r_h^{d-2}\left(1+\frac{r_h^2}{L^2}+\frac{q^2}{r_h^{2d-4}}\right)
\label{eq2.21}
\end{equation}
The black hole entropy and Hawking temperature can then be obtained from the horizon geometry and the surface gravity, respectively
\begin{equation}
T=\frac{d-2}{4\pi r_h}\left(1+\frac{d}{d-2}\frac{r_h^2}{L^2}-\frac{q^2}{r_h^{2d-4}}\right),\qquad S=\frac{\Omega_{d-1}r_h^{d-1}}{4G_N}
\end{equation}

\subsection{Extended boundary thermodynamics with central charge}
In the holographic framework, variations of the bulk cosmological constant and Newton's constant are associated with variations in the central charge of the dual conformal field theory (CFT) \cite{cft2,Visser}. For Einstein gravity, the central charge is expressed as
\begin{equation}
C=\frac{\Omega_{d-1}L^{d-1}}{16\pi G_N}
\label{eq2.23}
\end{equation}
The spatial volume of the boundary CFT can be varied independently by introducing a boundary curvature radius $R$, such that
\begin{equation}
\mathcal{V}=\Omega_{d-1}R^{d-1}
\label{eq2.24}
\end{equation}
With these identifications, the holographic dictionary relating the bulk and boundary thermodynamic quantities takes the form
\begin{equation}
S=\frac{A}{4G_N},\qquad E=M\frac{L}{R},\qquad T=\frac{\kappa}{2\pi}\frac{L}{R},\qquad \widetilde{\Phi}=\frac{\Phi}{R},\qquad \widetilde{Q}=QL
\label{eq2.25}
\end{equation}
By employing the Smarr relation to eliminate $\Theta$ and expressing the variations of $\Lambda$ in terms of $L$, the bulk first law can be recast as
\begin{align}
d\left(M\frac{L}{R}\right)
={}&
\frac{\kappa}{2\pi}\frac{L}{R}d\left(\frac{A}{4G_N}\right)+\frac{\Phi}{R}d(QL)-\frac{M}{d-1}\frac{L}{R}\frac{dR^{d-1}}{R^{d-1}}\nonumber\\
&+
\left(M\frac{L}{R}-\frac{\kappa A}{8\pi G_N}\frac{L}{R}-\frac{\Phi}{R}QL\right)\frac{d(L^{d-1}/G_N)}{L^{d-1}/G_N}
\label{eq2.26}
\end{align}
The bulk thermodynamic relations map onto the extended first law of the dual CFT,
\begin{equation}
dE = T\,dS + \widetilde{\Phi}\,d\widetilde{Q}-p\,d\mathcal{V}+\mu\,dC
\label{eq2.27}
\end{equation}
A comparison with the holographic first law yields the CFT pressure and the chemical potential conjugate to the central charge as
\begin{equation}
p=\frac{1}{d-1}\frac{E}{\mathcal{V}},\qquad\mu=\frac{1}{C}\left(E-TS-\widetilde{\Phi}\widetilde{Q}\right)
\label{eq2.28}
\end{equation}

To express the thermodynamic quantities in a convenient dimensionless form, we introduce
\begin{equation}
x=\frac{r_h}{L},
\qquad
y=\frac{q}{L^{d-2}}
\label{eq2.29}
\end{equation}
In terms of these variables, the thermodynamic quantities of the dual CFT are given by
\begin{equation}
S=4\pi Cx^{d-1},\qquad\widetilde{Q}=2\alpha(d-1)Cy,\qquad\widetilde{\Phi}=\frac{1}{\alpha R}\frac{y}{x^{d-2}}
\label{eq2.30}
\end{equation}
while the energy and temperature take the forms
\begin{align}
E&=\frac{d-1}{R}Cx^{d-2}\left(1+x^2+\frac{y^2}{x^{2d-4}}\right)
\label{eq2.31}\\
T&=\frac{d-2}{4\pi R x}\left(1+\frac{d}{d-2}x^2-\frac{y^2}{x^{2d-4}}\right)
\label{eq2.32}
\end{align}
The chemical potential associated with variations in the central charge is consequently
\begin{equation}
\mu=\frac{x^{d-2}}{R}\left(1-x^2-\frac{y^2}{x^{2d-4}}\right)
\label{eq2.33}
\end{equation}

Equations~\ref{eq2.30} - \ref{eq2.33} provide the thermodynamic relations required for the subsequent analysis. They serve as the starting point for constructing the off-shell free energy landscape and investigating the phase structure critical slowing down behaviour of the dual CFT side.

In the CFT description, besides the usual $(T,S)$ pair, we have three pairs of conjugate thermodynamic variables - $(\tilde{\Phi},\tilde{Q})$, $(p,\mathcal{V})$ and $(\mu,C)$ yielding $2^3=8$ ensembles. In this manuscript, however, we work in the fixed $(\tilde{Q},\mathcal{V},C)$ ensemble where a first order phase transition is observed \cite{Visser2}. We now write down the generalised free energy in the CFT side using $G=E-TS$, which gives
\begin{equation}
G=\frac{x^2 \tilde{Q}^2 L^d (L x)^{-d}}{4 \alpha^2 C (d-1) R}+\frac{C (d-1) \left(x^2+1\right) L^{-d} (L x)^d}{R x^2}-T\left( 4 \pi  C x^{d-1}\right)
\label{eq2.34a}
\end{equation}
In the fixed $(\tilde{Q},\mathcal{V},C)$ ensemble, the system exhibits a van der Waals-like phase transition, and is associated with the emergence of a critical point. The corresponding critical parameters can be determined by imposing the following conditions \begin{equation}
G^{1}=G^{3}=G^{3}=0
\label{eq2.34}
\end{equation}
In four spacetime dimension, the critical values come out to be 
\begin{equation}
x_c = 0.57735,\quad \tilde{Q}_c=0.59628,\quad T_c=0.44106
\label{eq2.36}
\end{equation}

The spinodal points can be calculated by imposing the condition that the first and second derivative of the generalised free energy with respect to the order parameter goes to zero simultaneously. Thus, the generalized free energy provides a unified framework for characterizing the phase structure, locating critical and spinodal points, and investigating the thermodynamic behaviour of the system.

The free energy landscape provides a useful framework for exploring the dynamical processes underlying black hole phase transitions. Its kinetic formulation is closely connected to the time-dependent Ginzburg - Landau theory \cite{Hohenberg}, in which the evolution of a system near criticality is described by the dissipative dynamics of a nonconserved order parameter \cite{model1,model2}.

In this approach, transitions between distinct black hole configurations are modeled as stochastic processes occurring over the underlying free energy landscape. Identifying the black hole entropy $S$ as the dynamical order parameter, its time evolution can be described by the Langevin equation
\begin{equation}
\frac{d^2 x}{dt^2}=-\zeta\frac{dx}{dt}-\frac{\partial G}{\partial x}+\eta(t)
\label{eq2.37}
\end{equation}
where $G$ denotes the generalized free energy, $\zeta$ is the dissipation coefficient, and $\eta(t)$ represents the stochastic force arising from thermal fluctuations. The dissipation coefficient characterizes the damping of the order-parameter dynamics due to its interaction with the surrounding thermal environment. The stochastic force is assumed to be Gaussian white noise with zero mean and a correlation function satisfying the fluctuation--dissipation relation,
\begin{equation}
\langle \eta(t)\rangle=0,\qquad\langle\eta(t)\eta(x)\rangle=2\zeta T\,\delta(t-x)
\label{eq2.38}
\end{equation}

Within this description, the generalized free energy determines the deterministic force acting on the order parameter, whereas dissipation and thermal noise govern its nonequilibrium evolution. The interplay between these contributions provides a stochastic description of transitions between black hole states and offers a framework for investigating the dynamical behaviour of the system near criticality.

\section{Critical Slowing Down in the CFT side}\label{section3}
\subsection{Analytical Description}

Critical slowing down is a characteristic feature of systems approaching a continuous phase transition. As the critical point is approached, the characteristic relaxation time increases significantly and, in the thermodynamic limit, may diverge. This phenomenon is generally accompanied by enhanced fluctuations and a growing temporal correlation length. In conventional critical phenomena, the relaxation time typically exhibits a power-law dependence on the reduced temperature,
\begin{equation}
\tau \sim |\epsilon|^{-\Delta},\qquad\epsilon=\frac{T-T_c}{T_c}
\label{eq:relaxation_scaling}
\end{equation}
where $\Delta$ denotes the dynamic critical exponent. A similar phenomenon can be investigated in black hole thermodynamics by examining the relaxation dynamics of the order parameter on the generalized free energy landscape.

To establish the dynamical description, let $x$ denote the order parameter of the system and $G(x)$ its generalized free energy. In the overdamped regime, the deterministic evolution of the order parameter is governed by the gradient-flow equation
\begin{equation}
\zeta\frac{dx}{dt}=-\frac{\partial G}{\partial x}
\label{eq:deterministic_dynamics}
\end{equation}
where $\zeta$ is the phenomenological dissipation coefficient. The equilibrium configurations correspond to stationary points of the generalized free energy, satisfying
\begin{equation}
\left.\frac{\partial G}{\partial x}\right|_{x=x_e}=0
\end{equation}
where $x_e$ denotes the equilibrium value of the order parameter.

The relaxation behaviour can be understood by expanding the generalized free energy around a stationary configuration. Retaining the leading nonvanishing terms gives
\begin{equation}
G(x)=G(x_e)+\frac{1}{2}G^{(2)}(x_e)(x-x_e)^2+\cdots
\label{eq:free_energy_expansion}
\end{equation}
where $G^{(n)}$ denotes the $n$th derivative with respect to $x$. Introducing a small deviation $\delta x=x-x_e$, Eq.~\eqref{eq:deterministic_dynamics} reduces to
\begin{equation}
\frac{d(\delta x)}{dt}=-\frac{G^{(2)}(x_e)}{\zeta}\delta x
\end{equation}
Its solution is an exponential relaxation,
\begin{equation}
\delta x(t)=\delta x(0)e^{-t/\tau},\qquad\tau=\frac{\zeta}{G^{(2)}(x_e)}
\label{eq:relaxation_time}
\end{equation}
The relaxation time is therefore controlled by the curvature of the generalized free energy at the equilibrium configuration. A flatter free energy landscape corresponds to a slower return towards equilibrium.

The situation changes qualitatively at the critical point, where the quadratic contribution to the free energy expansion vanishes. For a critical point characterized by
\begin{equation}
G^{(1)}(x_c)=G^{(2)}(x_c)=G^{(3)}(x_c)=0,\qquad G^{(4)}(x_c)\neq 0
\label{eq:critical_conditions}
\end{equation}
the leading contribution to the free energy expansion is quartic:
\begin{equation}
G(x)=G(x_c)+\frac{G^{(4)}(x_c)}{4!}(x-x_c)^4+\cdots
\label{eq:critical_free_energy}
\end{equation}
Consequently, the deterministic evolution equation becomes nonlinear,
\begin{equation}
\frac{d(\delta x)}{dt}=-\frac{G^{(4)}(x_c)}{6\zeta}(\delta x)^3,\qquad\delta x=x-x_c
\end{equation}
Integrating this equation yields
\begin{equation}
\delta x(t)=\frac{\delta x(0)}{\sqrt{1+\beta[\delta x(0)]^2t}},\qquad\beta=\frac{^{(4)}(x_c)}{3\zeta}
\label{eq:critical_algebraic_relaxation}
\end{equation}
Unlike the exponential relaxation away from criticality, the order parameter now approaches its equilibrium value algebraically. This change in the relaxation profile reflects the vanishing curvature of the generalized free energy landscape and provides a dynamical signature of criticality.

To incorporate thermal fluctuations, the deterministic description can be generalized to a stochastic Langevin equation,
\begin{equation}
\zeta\frac{dx}{dt}=-\frac{\partial G}{\partial x}+\eta(t)
\label{eq:stochastic_langevin}
\end{equation}
where $\eta(t)$ represents the thermal noise. Assuming Gaussian white noise, its statistical properties are specified by
\begin{equation}
\langle\eta(t)\rangle=0,\qquad\langle\eta(t)\eta(s)\rangle=2\zeta T\delta(t-s)
\label{eq:stochastic_noise}
\end{equation}

The stochastic dynamics can be studied by considering small fluctuations around a stable equilibrium configuration. Linearizing Eq.~\eqref{eq:stochastic_langevin} about $x_e$ gives
\begin{equation}
\frac{d(\delta x)}{dt}=-\frac{G^{(2)}(x_e)}{\zeta}\delta x+\frac{\eta(t)}{\zeta}
\label{eq:linearized_langevin}
\end{equation}
The formal solution of this equation is
\begin{equation}
\delta x(t)=\delta x(0)e^{-t/\tau}+\frac{1}{\zeta}\int_0^te^{-(t-s)/\tau}\eta(s)\,ds
\label{eq:formal_stochastic_solution}
\end{equation}
where $\tau$ is the relaxation time defined in Eq.~\eqref{eq:relaxation_time}.

An important quantity characterizing the stochastic evolution is the autocorrelation function of the order parameter,
\begin{equation}
C_x(t)=\left\langle\delta x(t)\delta x(0)\right\rangle
\label{eq:autocorrelation}
\end{equation}
For a stationary linear stochastic process, the autocorrelation function takes the exponential form
\begin{equation}
C_x(t)=C_x(0)e^{-t/\tau}
\label{eq:autocorrelation_exponential}
\end{equation}
with
\begin{equation}
\tau=\frac{\zeta}{G^{(2)}(x_e)}
\end{equation}
Thus, the autocorrelation time is directly determined by the curvature of the generalized free energy landscape. As the system approaches criticality, this curvature progressively decreases, resulting in an enhancement of the autocorrelation time. In the critical limit, where $G^{(2)}(x_c)=0$, the relaxation time diverges within the linearized description. This divergence provides a dynamical characterization of critical slowing down and establishes a connection between the geometric structure of the generalized free energy landscape and the temporal response of the system.

A comparable phenomenon is observed in the vicinity of the spinodal point, where the generalized free-energy landscape exhibits characteristics similar to those encountered near the critical point. In this regime, the conventional quadratic expansion of the free energy around equilibrium becomes inadequate, as its first and second derivatives with respect to the order parameter vanish. Consequently, the restoring force weakens, leading to a substantial enhancement of the autocorrelation time $\tau$. As the spinodal point is approached, $\tau$ grows without bound, indicating a pronounced slowing down of the underlying stochastic dynamics.
\begin{figure}[h!]
    \centering
    \begin{subfigure}[b]{0.45\textwidth}
        \centering
        \includegraphics[width=\textwidth]{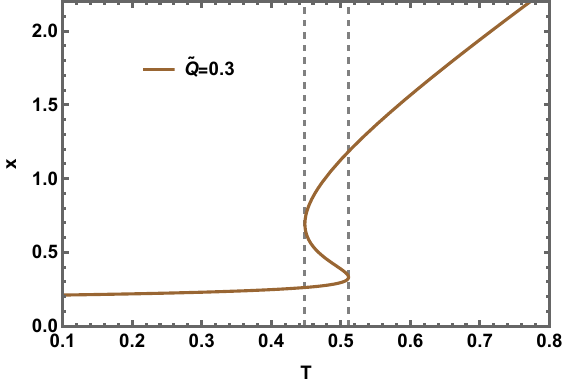}
        \caption{}
        \label{f1a}
    \end{subfigure}
    \hspace{0.8cm}
    \begin{subfigure}[b]{0.45\textwidth}
        \centering
        \includegraphics[width=\textwidth]{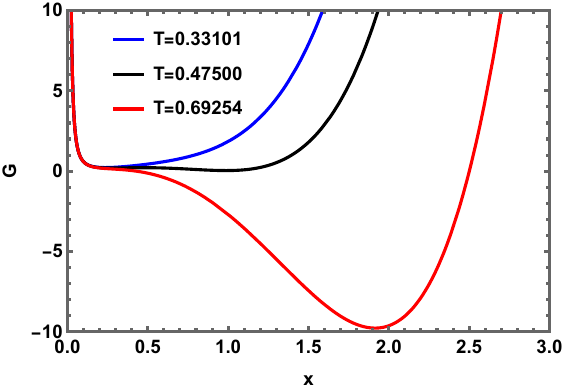}
        \caption{}
        \label{f1b}
    \end{subfigure}

    \caption{The left panel shows the temperature profile of the RN-AdS black hole in the dual CFT description for fixed $(\tilde{Q},\mathcal{V},C)$ ensemble for $\tilde{Q}<\tilde{Q}_c$. Here the dashed lines denote the two spinodal temperatures. The right panel shows the free energy landscapes for different values of the Hawking temperature.}
    \label{f1}
\end{figure}

Figure~\ref{f1a} illustrates the temperature profile of the RN AdS black hole in the dual CFT description. The corresponding free energy landscapes evaluated at the two spinodal temperatures are presented in Fig.~\ref{f1b}. The blue and red curves represent the landscapes at the left and right spinodal temperatures, respectively, while the black curve corresponds to an intermediate temperature between the two spinodals. As the ensemble temperature varies across this interval, the free energy landscape undergoes a qualitative transformation, characterized by the successive formation and disappearance of its extrema.
\begin{figure}[h!]
    \centering
    \begin{subfigure}[b]{0.45\textwidth}
        \centering
        \includegraphics[width=\textwidth]{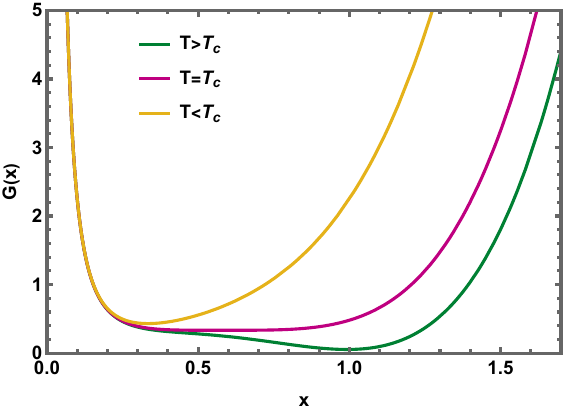}
        \caption{}
        \label{f2a}
    \end{subfigure}
    \hspace{0.8cm}
    \begin{subfigure}[b]{0.45\textwidth}
        \centering
        \includegraphics[width=\textwidth]{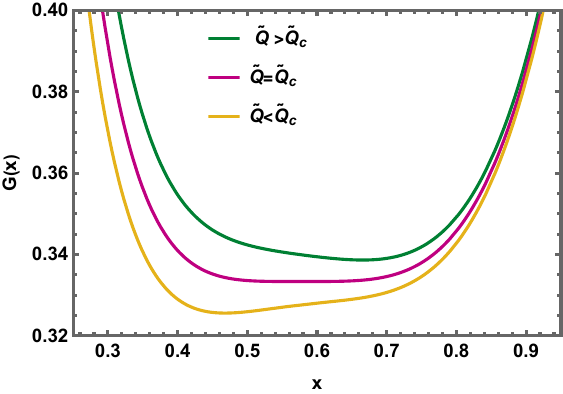}
        \caption{}
        \label{f2b}
    \end{subfigure}

    \caption{CFT free energy landscapes for varying ensemble temperature (left) and CFT charge $\tilde{Q}$ (right)}
    \label{f2}
\end{figure}

Figure~\ref{f2} further depicts the evolution of the generalized free energy landscape with variations in the ensemble temperature $T$ and the CFT charge $\widetilde{Q}$. A notable feature is the pronounced flattening of the landscape as the system approaches criticality as indicated by the magenta curve. The flattening of the free energy landscape reduces the effective restoring force acting on fluctuations around equilibrium, thereby contributing to the onset of critical slowing down. In contrast, sufficiently far from the critical point, the free energy profile recovers a well-defined single minimum structure as the temperature or charge is varied away from its respective critical value.

\subsection{Numerical Results}
We next examine the stochastic evolution of the system through numerical simulations of the Langevin equation governing the dynamics of the order parameter on the generalized free energy landscape. Using the resulting stochastic trajectories, we evaluate the autocorrelation function and determine the associated autocorrelation time by fitting its temporal decay to an exponential profile. Alongside the autocorrelation time, we investigate the variance of the order-parameter fluctuations and calculate the smallest nonzero eigenvalue of the corresponding Fokker - Planck operator to obtain an additional characterization of the relaxation dynamics. While the variance quantifies the magnitude of fluctuations around equilibrium, the smallest nonzero Fokker - Planck eigenvalue provides an estimate of the slowest characteristic relaxation rate of the system. For simplicity, the dissipation coefficient is fixed at $\zeta=1$ throughout the numerical analysis. Further details of the numerical implementation and computational procedure are provided in Appendix \ref{appendix A}.
\begin{figure}[h!]
\centering
\includegraphics[width=0.97\textwidth]{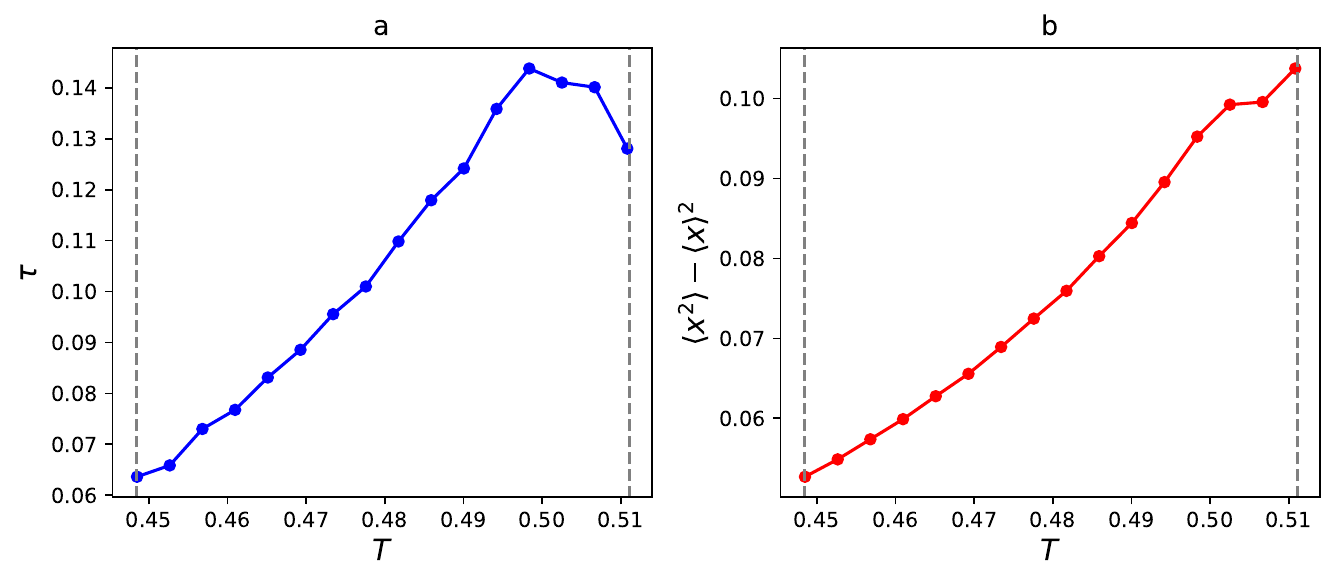}
\caption{Correlation time $\tau$ and variance $\langle x^2\rangle-\langle x\rangle^2$ of trajectories as a function of $T$ near small black hole.}
\label{f3}
\end{figure}

We begin by examining the stochastic dynamics in the neighbourhood of the spinodal points. As the ensemble temperature increases from the left spinodal towards the right spinodal, the free energy minimum associated with the small black hole configuration progressively becomes less pronounced and ultimately disappears as illustrated in Fig.~\ref{f1b}, where the blue and red curves represent the free energy landscapes at the left and right spinodal temperatures, respectively. The progressive flattening of the potential well reduces the effective restoring force acting on the order parameter, thereby slowing its relaxation towards equilibrium. Consequently, both the autocorrelation time and the variance of the stochastic trajectories associated with the small black hole phase exhibit a substantial increase with temperature, as demonstrated in the left panels of Fig.~\ref{f3}. This enhancement of the relaxation timescale and fluctuations provides a characteristic signature of critical slowing down.
\begin{figure}[h!]
\centering
\includegraphics[width=0.97\textwidth]{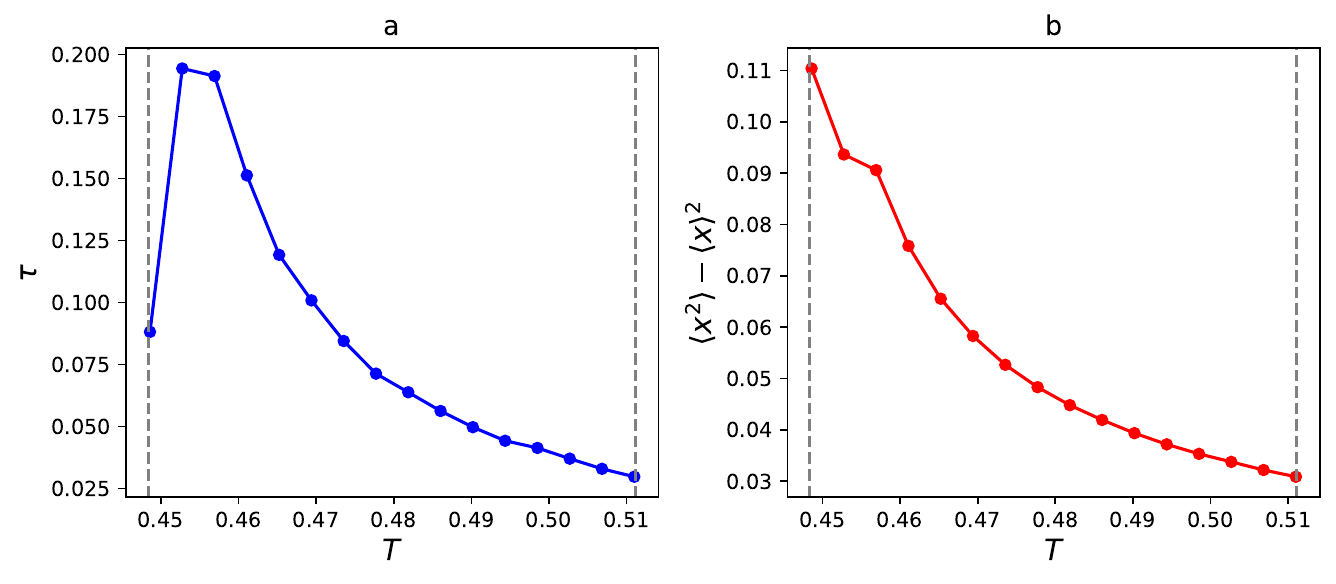}
\caption{Correlation time $\tau$ and variance $\langle x^2\rangle-\langle x\rangle^2$ of trajectories as a function of $T$ near large black hole.}
\label{f4}
\end{figure}

A complementary behaviour is observed when the ensemble temperature is decreased from the right spinodal towards the left spinodal. During this evolution, the free energy minimum corresponding to the large black hole configuration gradually becomes shallower before eventually disappearing, as illustrated by the transition from the red curve to the blue curve in Fig.~\ref{f1b}. The resulting reduction in the local curvature of the free energy landscape weakens the restoring force and leads to progressively slower relaxation. Accordingly, the autocorrelation time and the variance of the stochastic trajectories associated with the large black hole phase increase as the temperature decreases, as shown in Fig.~\ref{f4}. These observations provide further evidence of critical slowing down and demonstrate that the dynamical response near the spinodal boundaries depends on the thermodynamic branch under consideration.
\begin{figure}[t]
\centering
\includegraphics[width=0.97\textwidth]{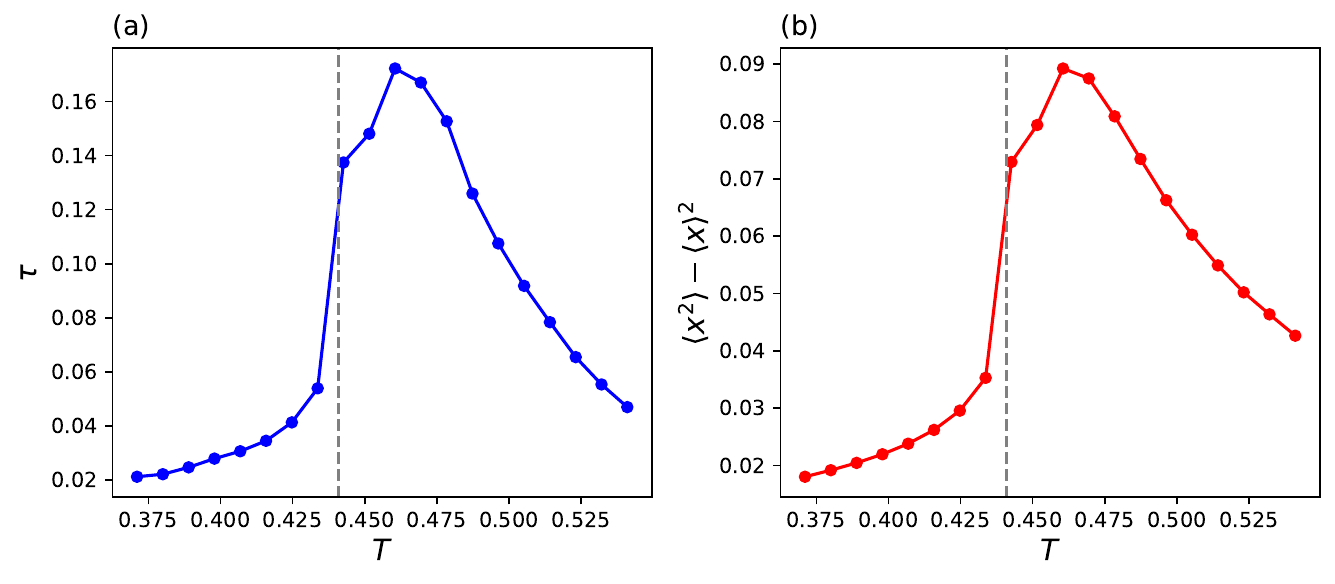}
\caption{Correlation time $\tau$ and variance $\langle x^2\rangle-\langle x\rangle^2$ of trajectories as a function of $T$. The grey dashed line represents the critical temperature.}
\label{f5}
\end{figure}

\begin{figure}[h!]
\centering
\includegraphics[width=0.97\textwidth]{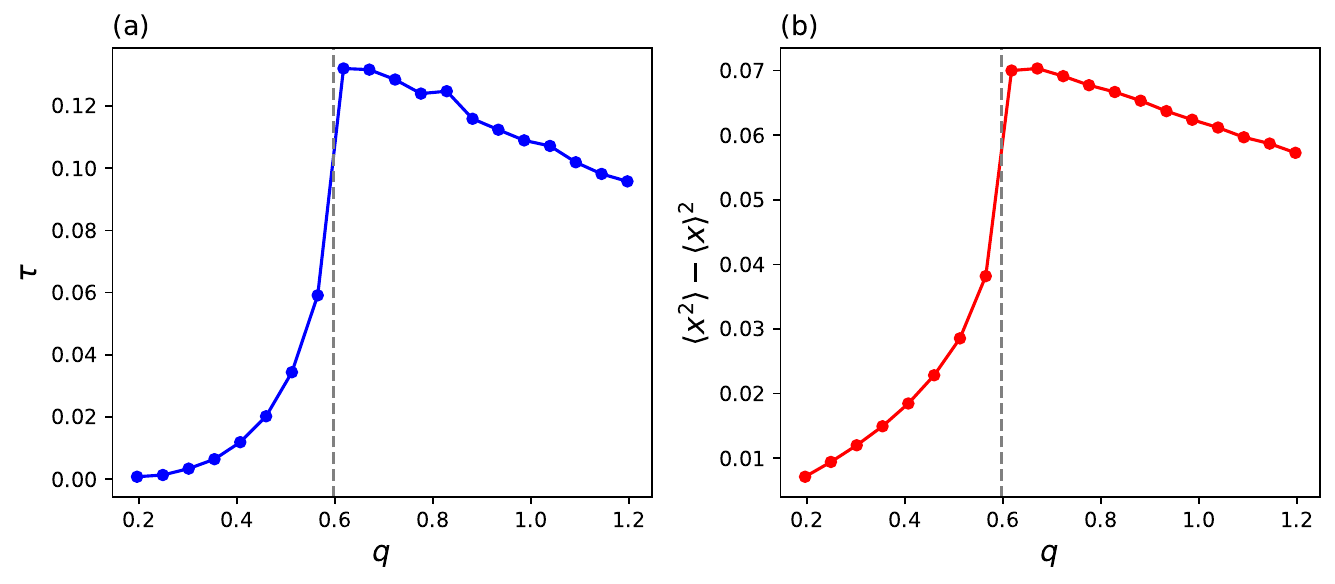}
\caption{Correlation time $\tau$ and variance $\langle x^2\rangle-\langle x\rangle^2$ of trajectories as a function of $\tilde{Q}$. The grey dashed line represents the critical CFT charge.}
\label{f6}
\end{figure}

Next we turn our focus in the vicinity of the critical point where we simulate the Langevin equation around the critical point of temperature and charge and see the behaviour of the autocorrelation time and its variance. For this purpose, we investigate along two distinct thermodynamic paths. In the first one, we fix the CFT charge at its critical value $\tilde{Q}=\tilde{Q}_c$ and vary the temperature around its critical value $T_c$. For the second case, the temperature is kept fixed at it critical value while the charge is varied around the critical value. The evolution of the autocorrelation time and variance along the corresponding trajectories is presented in Figs.~\ref{f5} and \ref{f6}, respectively.

We observe that as the system approaches criticality, both the autocorrelation time $\tau$ and variance increases significantly and then decreases as we move away from criticality. This behaviour suggests the the flattening of the free energy landscape as criticality is reached as has been demonstrated in Fig \ref{f2}.
\begin{figure}[h!]
    \centering
    \begin{subfigure}[b]{0.45\textwidth}
        \centering
        \includegraphics[width=\textwidth]{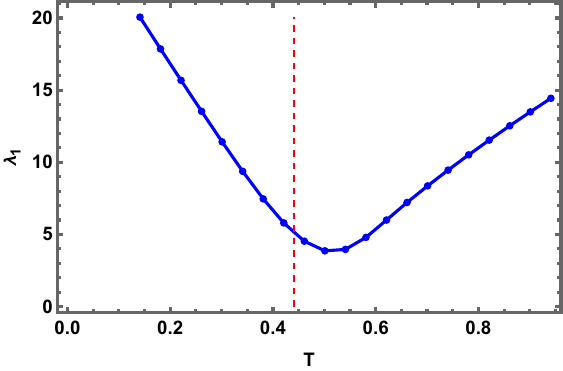}
        \caption{}
        \label{f7a}
    \end{subfigure}
    \hspace{0.8cm}
    \begin{subfigure}[b]{0.44\textwidth}
        \centering
        \includegraphics[width=\textwidth]{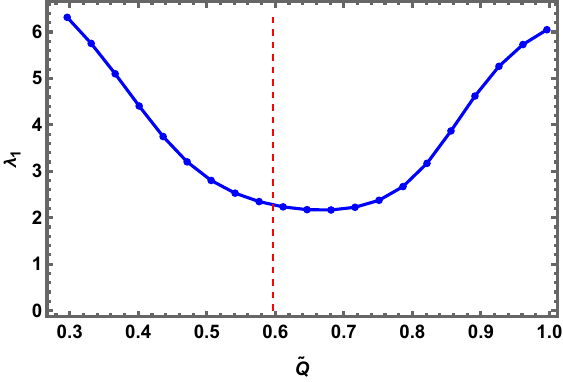}
        \caption{}
        \label{f7b}
    \end{subfigure}

    \caption{Smallest non-zero eigenvalue of the Fokker-Planck equation with (a) temperature and (b) CFT charge}
    \label{f7}
\end{figure}

We further characterize the relaxation dynamics by examining the smallest nonzero eigenvalue of the associated Fokker-Planck operator. As illustrated in Fig.~\ref{f7}, this eigenvalue undergoes a pronounced suppression in the vicinity of the critical point. This behaviour is consistent with the enhancement of the autocorrelation time and the variance of the stochastic trajectories discussed above, providing an independent indication of the emergence of critical slowing down. The numerical methodology employed to determine the Fokker-Planck eigenvalues is described in Appendix~\ref{appendix A}.

An interesting feature of the results is that the slowest relaxation does not necessarily occur precisely at the critical point. Along both thermodynamic paths, corresponding to fixed angular momentum and fixed temperature, the strongest suppression of the lowest nonzero eigenvalue is observed slightly below the respective critical values. A similar displacement of the minimum eigenvalue from the critical point has also been reported in previous investigations as well \cite{slowing,Mozib}.

\section{Dynamical Scaling and Universality}\label{section4}
\subsection{Numerical Extraction of Delta}

The holographic correspondence provides an avenue for investigating the dynamical manifestations of black hole critical phenomena in the dual conformal field theory. In particular, the thermodynamic critical behaviour of the bulk system can be translated into the boundary theory, allowing us to examine how the relaxation dynamics of the CFT respond to the approach towards criticality. An important question in this context is whether the critical slowing down observed in the bulk thermodynamic description is accompanied by a universal scaling behaviour in the dual CFT. To address this, we investigate the scaling of the autocorrelation time near the critical point and extract the associated dynamical critical exponent. This enables us to examine the universality of the relaxation dynamics and explore its dependence on the underlying thermodynamic system.

The autocorrelation time is expected to exhibit a power-law divergence near criticality of the form
\begin{equation}
\tau \sim |\epsilon|^{-\Delta}
\label{eq4.1}
\end{equation}
where $\epsilon$ denotes the reduced temperature or charge, defined accordingly along different thermodynamic paths as
\begin{equation}
\epsilon=\frac{T-T_c}{T_c}, \quad \epsilon_Q=\frac{\tilde{Q}-\tilde{Q}_c}{\tilde{Q}_c}
\label{eq4.2}
\end{equation}
and $\Delta$ is the dynamical critical exponent governing the divergence of the relaxation time. In our previous work \cite{Mozib}, we demonstrated that the dynamical critical exponent $\Delta$ takes a value close to $0.66$, consistent with the analytical prediction $\Delta=2/3$. This behaviour was established through an investigation of three distinct black hole configurations, namely, the RN-AdS, Kerr-AdS, and Bardeen black holes, within the bulk gravitational framework. Furthermore, an analytical justification for this scaling exponent was obtained by examining the expansion of the generalized free energy near criticality. Motivated by these findings, the present work extends the investigation to the holographically dual CFT description, where we aim to extract the corresponding dynamical critical exponent and examine whether the scaling behaviour identified in the bulk is preserved in the boundary theory.
\begin{figure}[h!]
\centering
\includegraphics[width=0.97\textwidth]{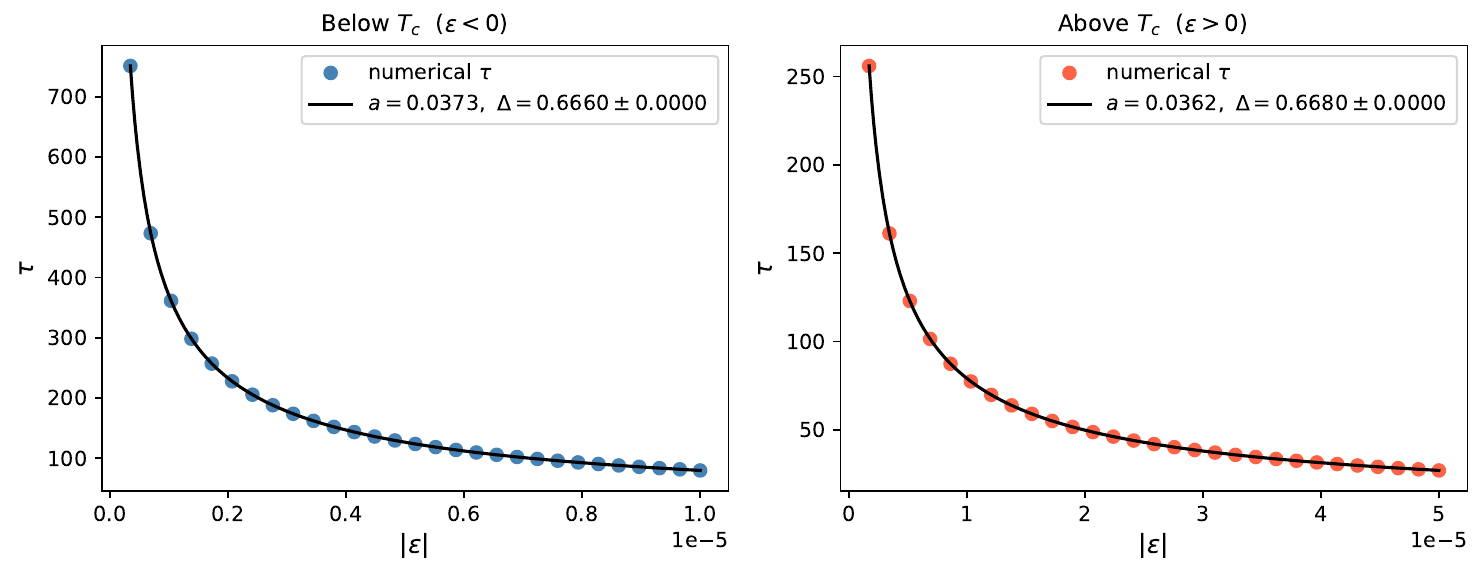}
\caption{Power-law scaling of the autocorrelation time $\tau$ with the reduced temperature $|\epsilon|$ on both sides of the critical point.}
\label{f8}
\end{figure}

\begin{figure}[h!]
\centering
\includegraphics[width=0.97\textwidth]{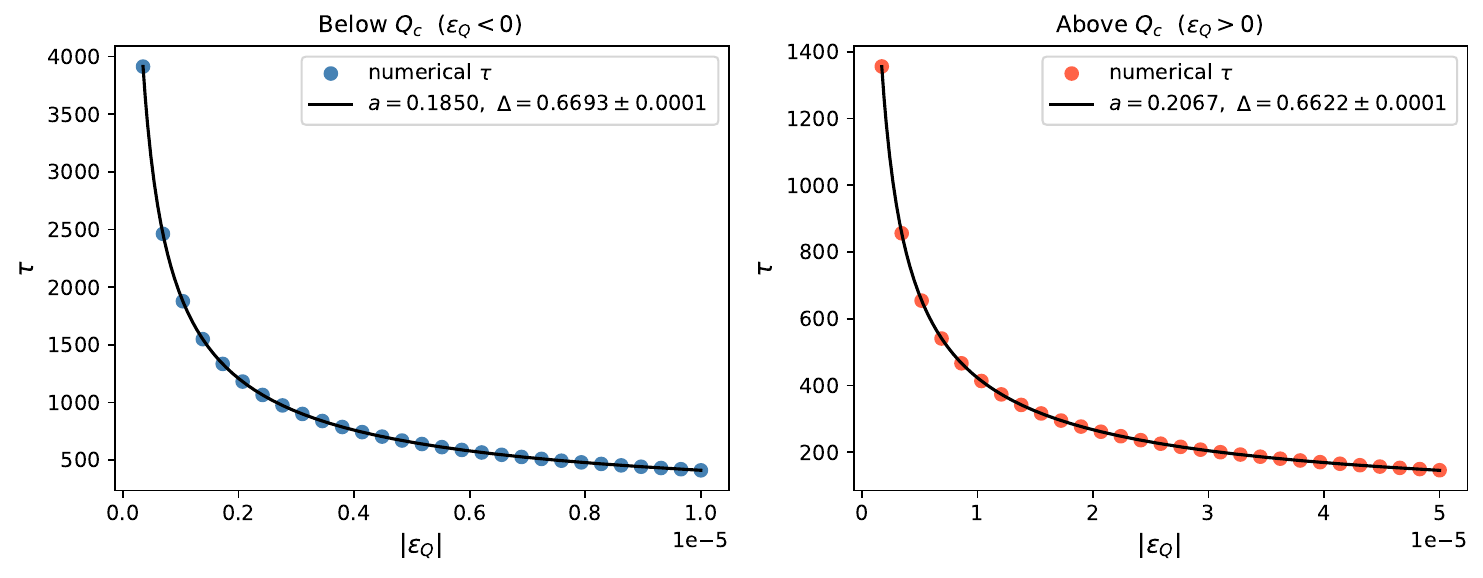}
\caption{Power-law scaling of the autocorrelation time $\tau$ with $|\epsilon|$ on both sides of the critical point.}
\label{f9}
\end{figure}

To extract the dynamical critical exponent, we perform a power-law fit to the numerically evaluated autocorrelation time in the vicinity of the critical point, using the scaling relation
\begin{equation}
\tau=a|\epsilon|^{-\Delta}
\label{eq4.3}
\end{equation}
where $a$ and $\Delta$ are treated as fitting parameters.
In Fig.~\ref{f8}, we display the numerical results for the autocorrelation time along with the corresponding power-law fits near the critical point. We consider two distinct thermodynamic trajectories: varying the temperature while keeping the CFT charge fixed (Fig.~\ref{f8}), and varying the CFT charge at a fixed temperature (Fig.~\ref{f9}). For both trajectories, the extracted dynamical critical exponent is found to be approximately $\Delta\simeq 0.66$, demonstrating a consistent scaling behaviour in the vicinity of criticality.

Although the scaling exponent remains nearly unchanged, the prefactor $a$ exhibits a dependence on both the thermodynamic trajectory and the direction of approach to the critical point, corresponding to $\epsilon<0$ and $\epsilon>0$. Furthermore, the autocorrelation time grows rapidly as $\epsilon\rightarrow 0$, reflecting the increasingly slow relaxation of the system near criticality. Notably, the numerically extracted exponent is in close agreement with the analytical prediction $\Delta=2/3$. We provide a detailed analytical derivation of this exponent based on the critical expansion of the generalized free energy landscape in the Appendix of \cite{Mozib}.

These results indicate that the dynamical scaling behaviour previously identified in the bulk gravitational description \cite{Mozib} is also manifested in the dual CFT. In particular, the close agreement between the numerically extracted exponent and the analytical prediction $\Delta=2/3$ suggests that the critical relaxation dynamics are insensitive to the choice of thermodynamic trajectory. From the holographic perspective, this provides evidence that the characteristic scaling of critical slowing down is retained under the bulk-boundary correspondence. Thus, the dual CFT offers an alternative framework for investigating the universal dynamical features associated with black hole critical phenomena.

\subsection{Two Parameter Scaling with Temperature and Central Charge}

The results presented in the preceding subsection demonstrate that the autocorrelation time exhibits a consistent power-law divergence near the critical point, with the dynamical critical exponent taking the value $\Delta=2/3$ along both thermodynamic trajectories considered. We now extend this analysis by treating the central charge $C$ of the dual conformal field theory as an independent control parameter. This allows us to investigate whether the critical scaling behaviour persists when temperature and central charge are varied simultaneously, rather than along separate one-dimensional paths.

On the CFT side, the generalized free-energy landscape is given by

\begin{equation}
G(x,T,\tilde{Q},C)=C\left(3x^2-4\pi T x^3+3x^4\right)+\frac{\tilde{Q}^2}{16Cx^2}
\label{eq4.4}
\end{equation}

This is just a rearrangement of the terms of Eq. \ref{eq2.34a} after setting $L=R=1$. Writing $q\equiv\tilde{Q}/C$, Eq. \ref{eq4.4} becomes $G=C\tilde{G}(x,T,q)$ The equilibrium configurations are determined by the stationary condition

\begin{equation}
G'(x_e)=0
\end{equation}

where the prime denotes differentiation with respect to $x$. At the critical point, the first three derivatives of the generalized free energy vanish, while its fourth derivative remains finite and nonzero. For the present system, choosing $C=1$ and $\tilde{Q}=\tilde{Q}_*$, the criticality conditions yield

\begin{equation}
x_c=\frac{1}{\sqrt{3}},\qquad
T_c=\frac{4\sqrt{3}}{5\pi},\qquad
\tilde{Q}_*^2=\frac{16}{45}
\end{equation}
 which are simply numerically equal to the values obtained in Eq. \ref{eq2.36}. Also we have,
\begin{equation}
G^{(4)}(x_c)=144
\end{equation}

To examine the response of the system to simultaneous variations of temperature and central charge, we consider small deviations from the critical point. At fixed charge $\tilde{Q}$, the free energy depends on $\tilde{Q}$ and $C$ only through $q=\tilde{Q}/C$, so that criticality
requires $q=q_c\equiv 16/45$, i.e. $C_c=\tilde{Q}/q_c$, with $(x_c,T_c)$ independent of $C$. We fix $\tilde{Q}=\tilde{Q}_*\equiv q_c$, for which $C_c=1$, and define
\begin{equation}
  t=T-T_c,\qquad \delta C=C-C_c,\qquad \Delta x=x_e-x_c .
\end{equation}
For any other fixed $\tilde{Q}$ the analysis is unchanged, with $C_c=\tilde{Q}/q_c$.

Expanding the equilibrium condition $G'(x_e)=0$ around $(x_c,T_c,C=1)$, and using the vanishing of the first three derivatives of $G$ at criticality, we obtain, to leading order,

\begin{equation}
A_Tt+A_C\delta
+\frac{G^{(4)}(x_c)}{6}(\Delta x)^3
+\cdots=0
\end{equation}

where the coefficients associated with the temperature and central-charge variations are defined as

\begin{equation}
A_T=
\left.\frac{\partial G'}{\partial T}\right|_c
=-12\pi x_c=-4\pi,
\qquad
A_C=
\left.\frac{\partial G'}{\partial C}\right|_c
=\frac{q_*^2}{4x_c^3}
=\frac{4\sqrt{3}}{15}
\end{equation}

Here, the subscript $c$ indicates evaluation at the critical point. Both coefficients are nonzero, indicating that variations in temperature and central charge contribute to the leading-order displacement of the equilibrium configuration. It is therefore natural to introduce a single scaling field,

\begin{equation}
u\equiv A_Tt+A_C\delta
\end{equation}

In terms of this variable, the equilibrium condition reduces to

\begin{equation}
u+24(\Delta x)^3\simeq0
\end{equation}

which gives

\begin{equation}
\Delta x=-\left(\frac{u}{24}\right)^{1/3}
\end{equation}

Consequently, the curvature of the generalized free-energy landscape evaluated at equilibrium behaves as

\begin{equation}
G''(x_e)
\simeq \frac{G^{(4)}(x_c)}{2}(\Delta x)^2
=72\left(\frac{|u|}{24}\right)^{2/3}
\end{equation}

Within the overdamped Langevin description, the relaxation time is inversely proportional to the curvature of the free-energy landscape, with the dissipation coefficient $\zeta$ entering as

\begin{equation}
\tau=\frac{\zeta}{G''(x_e)}
\end{equation}

Combining the above expressions, we arrive at

\begin{equation}
\tau=
\frac{24^{2/3}\zeta}{72}|u|^{-2/3}
\simeq 0.1156\,\zeta\,|u|^{-2/3}
\end{equation}

This result shows that the critical relaxation time is governed by the same dynamical exponent $\Delta=2/3$, but with the distance from criticality characterized by the combined scaling field $u$, rather than by temperature or charge variations separately. In particular, along a pure-temperature trajectory with $C=1$, one has $u=A_T(T-T_c)$. Expressing the result in terms of the reduced temperature $\epsilon=(T-T_c)/T_c$, the relaxation time becomes

\begin{equation}
\tau=a|\epsilon|^{-2/3},
\qquad
a=
\frac{24^{2/3}\zeta}
{72(|A_T|T_c)^{2/3}}
\simeq 0.0369\,\zeta
\end{equation}

For $\zeta=1$, this analytical amplitude is in close agreement with the numerical values obtained in the preceding subsection.

An important consequence of the above result is that, to leading order, the relaxation time depends on the direction of approach to the critical point only through the value of $|u|$. Thus, different trajectories in the $(T,C)$ plane that correspond to the same value of $|u|$ exhibit the same leading-order relaxation time. This behaviour is expected to hold for generic approaches to the critical point, for which $u$ varies linearly with the distance from criticality. There is, however, a special direction along which the leading-order contribution vanishes, namely,

\begin{equation}
A_T\,dT+A_C\,dC=0,
\qquad
\left.\frac{dC}{dT}\right|_{u=0}
=-\frac{A_T}{A_C}
=\frac{15\pi}{\sqrt{3}}
\end{equation}
Along this direction, higher-order terms in the expansion become relevant, and the scaling behaviour requires a separate analysis.

To test the analytical prediction, we numerically evaluate the relaxation time using the full generalized free-energy landscape, without invoking the near-critical expansion. We consider twelve different trajectories passing through the critical point in the $(T,C)$ plane, parametrized as

\begin{equation}
T=T_c+s\cos\theta,\qquad
C=1+30s\sin\theta,\qquad
\theta=\frac{2\pi k}{12}
\end{equation}
where $k=0,1,\ldots,11$, and $s$ is logarithmically varied over the range $10^{-6}\leq s\leq 3\times10^{-3}$. The factor $30$ balances the sizes of $A_T$ and $A_C$. The numerical equilibrium radius is obtained by solving $G'(x_e)=0$, and the relaxation time is subsequently evaluated from the full free-energy landscape, with $\zeta=1$. We then plot the resulting relaxation times against the absolute value of the scaling field $|u|$.
\begin{figure}[h!]
\centering
\includegraphics[width=0.7\textwidth]{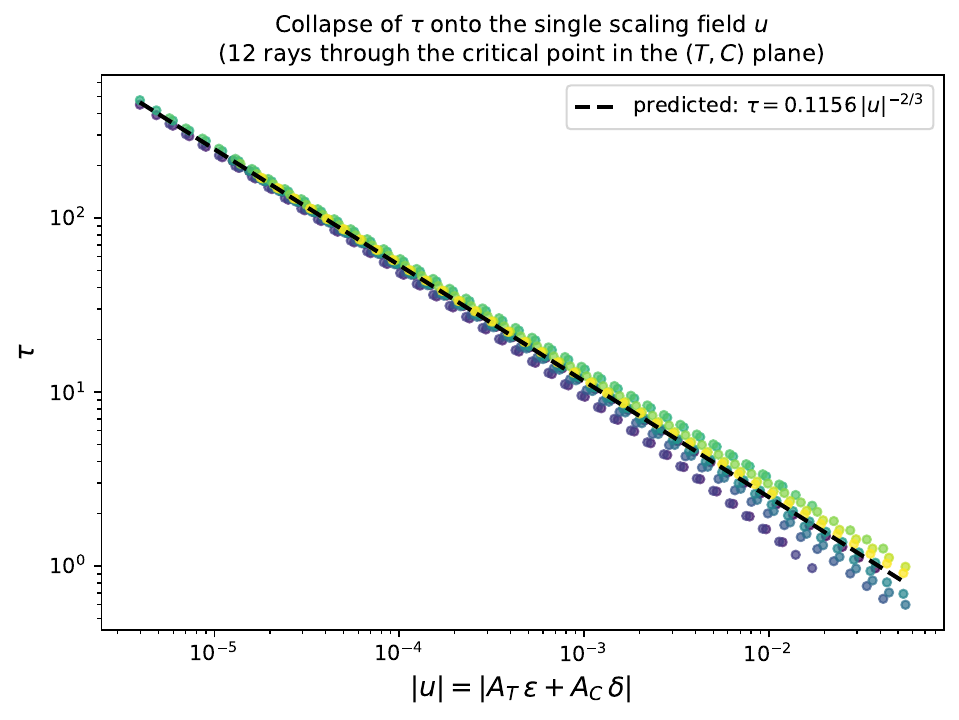}
\caption{Scaling of $\tau$ along twelve different trajectories in the $(T,C)$ plane. The numerical data, obtained from the full generalized free-energy landscape, collapse onto a single curve and agree with the analytical prediction $\tau=0.1156|u|^{-2/3}$ (dashed line)}
\label{f10}
\end{figure}

The results are presented in Fig.~\ref{f10}. Despite the different directions of approach, the numerical data exhibit a remarkable collapse onto a single curve over nearly four decades in $|u|$. Furthermore, the collapsed data agree closely with the analytical prediction, without requiring an additional fitting parameter. Small deviations become visible at larger values of $|u|$, where subleading contributions neglected in the near-critical expansion become increasingly important. These findings provide numerical evidence that the combined variable $u$ captures the leading critical behaviour of the relaxation dynamics when temperature and central charge are varied simultaneously.

\section{Relaxation and Scrambling in the Dual CFT}\label{section5}

\subsection{Motivation}

The critical slowing down discussed in the preceding sections is associated with the increasingly weak restoring force acting on fluctuations of the order parameter near the critical point. In the holographic description, however, the boundary theory also possesses a characteristic microscopic timescale associated with the scrambling of information among its degrees of freedom \cite{scrambling,scrambling2,scrambling3}. This raises a natural question: how does the relaxation timescale of the thermodynamic order parameter compare with the scrambling timescale of the dual CFT, and how does their relative hierarchy depend on the central charge?

The scrambling time provides a useful measure of the timescale over which initially localized information becomes distributed among the microscopic degrees of freedom of a strongly interacting quantum system. For holographic theories with a large number of degrees of freedom, it is conventionally estimated using the fast-scrambling relation. Comparing this timescale with the relaxation time obtained from the generalized free-energy landscape allows us to identify the regime in which the thermodynamic order parameter becomes slower than the underlying microscopic scrambling dynamics.

In what follows, we investigate this comparison along the critical locus of the CFT, where the critical temperature and the equilibrium value of the order parameter remain independent of the central charge. This provides a convenient setting for isolating the effect of \(C\) on the hierarchy between the two timescales.

\subsection{Relaxation time along the critical locus}

We begin by recalling that the CFT free-energy landscape can be expressed in terms of the rescaled charge \(q=\tilde{Q}/C\) as

\begin{equation}
G(x,T,\tilde{Q},C)=C\,\widetilde G(x,T, q)
\end{equation}

Consequently, the critical point remains unchanged when the central charge is varied while the rescaled charge is held at its critical value. We therefore consider the critical locus

\begin{equation}
\tilde{Q}=\tilde{Q}_c(C)=Cq_*
\end{equation}

where \(q_*\) denotes the critical charge at \(C=1\). Along this locus, the critical temperature and equilibrium radius are given by \(T_c\) and \(x_c\), respectively, and are independent of \(C\).

The relaxation time obtained from the linearized dynamics around the equilibrium configuration is

\begin{equation}
\tau=\frac{\zeta}{G''(x_e)}
\end{equation}

where \(\zeta\) is the dissipation coefficient and \(x_e\) denotes the equilibrium value of the order parameter. Since the free-energy landscape is proportional to \(C\) along the critical locus, its curvature scales linearly with the central charge. Using the critical scaling derived in the preceding section, we obtain

\begin{equation}
\tau(\epsilon,C)=\frac{a}{C}|\epsilon|^{-2/3}
\label{eq:tau_C}
\end{equation}

where \(\epsilon=(T-T_c)/T_c\) is the reduced temperature and

\begin{equation}
a=\frac{24^{2/3}\zeta}{72(|A_T|T_c)^{2/3}}
\simeq 0.0369\,\zeta
\end{equation}

Here, we have used the critical coefficients obtained from the near-critical expansion, and the numerical value corresponds to the present normalization of the free-energy landscape. Equation~\ref{eq:tau_C} shows that, at a fixed distance from criticality, increasing the central charge reduces the relaxation time as \(C^{-1}\). Nevertheless, the critical divergence persists for every finite \(C\), suggesting that sufficiently close to criticality the relaxation timescale can become comparable to, and eventually exceed, the microscopic scrambling timescale.

\subsection{Scrambling time in the dual CFT}

For a holographic theory with a large number of degrees of freedom, the scrambling time is estimated by the fast-scrambling relation

\begin{equation}
t_*=\frac{\beta}{2\pi}\ln S
=\frac{1}{2\pi T}\ln S
\end{equation}

where \(\beta=1/T\) and \(S\) denotes the entropy of the system. In the present analysis, we employ the area-law expression

\begin{equation}
S=C\pi x_e^2
\end{equation}

which makes explicit the extensive dependence of the entropy on the central charge. The scrambling time consequently takes the form

\begin{equation}
t_*=\frac{1}{2\pi T}\ln(C\pi x_e^2)
\label{eq:scrambling_time}
\end{equation}

An important feature of Eq.~\ref{eq:scrambling_time} is its logarithmic dependence on \(C\), in contrast to the inverse linear dependence of the relaxation time in Eq.~(\ref{eq:tau_C}). Thus, away from the critical point, the relaxation time becomes increasingly small relative to the scrambling time as \(C\) grows. The critical divergence of \(\tau\), however, can compensate for this suppression sufficiently close to \(T_c\).

We emphasize that the expression above is meaningful in the large-entropy regime, \(S\gg1\), where the holographic scrambling-time estimate is applicable. Accordingly, the comparison below is restricted to sufficiently large central charge.

\subsection{Crossover between relaxation and scrambling}

The competition between the two timescales can be characterized by defining a crossover reduced temperature \(\hat\epsilon\) through

\begin{equation}
\tau(\hat\epsilon,C)=t_*(\hat\epsilon,C)
\label{eq:crossover_condition}
\end{equation}

Substituting Eqs.~\ref{eq:tau_C} and \ref{eq:scrambling_time} into this condition gives

\begin{equation}
\frac{a}{C}\hat\epsilon^{-2/3}
=
\frac{1}{2\pi T}\ln(C\pi x_e^2)
\end{equation}

In the large-\(C\) regime, the crossover occurs very close to the critical point. We may therefore evaluate the slowly varying quantities \(T\) and \(x_e\) at their critical values, \(T_c\) and \(x_c\), respectively. Solving for the crossover reduced temperature, we obtain

\begin{equation}
\hat\epsilon=
\left[
\frac{2\pi T_c a}
{C\ln(C\pi x_c^2)}
\right]^{3/2}
\label{eq:epsilon_hat}
\end{equation}

Since \(x_c\) is independent of \(C\) along the critical locus, the asymptotic behaviour at large central charge becomes

\begin{equation}
\hat\epsilon
\underset{C\gg1}{\sim}
(2\pi T_c a)^{3/2}(C\ln C)^{-3/2}
\label{eq:epsilon_asymptotic}
\end{equation}

For the present critical point, \(x_c=1/\sqrt{3}\), and hence

\begin{equation}
\hat\epsilon=
\left[
\frac{2\pi T_c a}
{C\ln(C\pi/3)}
\right]^{3/2}
\end{equation}

with the asymptotic coefficient

\begin{equation}
(2\pi T_c a)^{3/2}\simeq 0.0327\,\zeta^{3/2}
\end{equation}

Equation~\ref{eq:epsilon_asymptotic} is the central result of this comparison. It shows that the region in which the relaxation time becomes comparable to or larger than the scrambling time shrinks rapidly as the central charge increases. In particular, the crossover scale is not described by a pure power law in \(C\); the logarithmic contribution inherited from the scrambling time produces an additional factor \((\ln C)^{-3/2}\).

The approximation leading to Eq.~\ref{eq:epsilon_hat} is self-consistent provided that \(\hat\epsilon\ll1\). As shown below, the crossover values obtained numerically lie well within this near-critical regime over the range of central charges considered.

\subsection{Numerical verification and physical implications}

To test the analytical prediction, we determine the crossover scale numerically using the full nonlinear free-energy landscape rather than the near-critical approximation. For each chosen value of \(C\), we move along the critical locus \(q=Cq_*\) and parameterize the temperature by

\begin{equation}
T=T_c(1-\epsilon)
\end{equation}

where \(\epsilon>0\) measures the distance below the critical temperature. At each trial value of \(\epsilon\), we solve the full equilibrium condition \(G'(x_e)=0\), retain the stable equilibrium solution, and evaluate the relaxation and scrambling times using

\begin{equation}
\tau=\frac{\zeta}{G''(x_e)},
\qquad
t_*=\frac{1}{2\pi T}\ln(C\pi x_e^2)
\end{equation}
\begin{figure}[h!]
\centering
\includegraphics[width=0.6\textwidth]{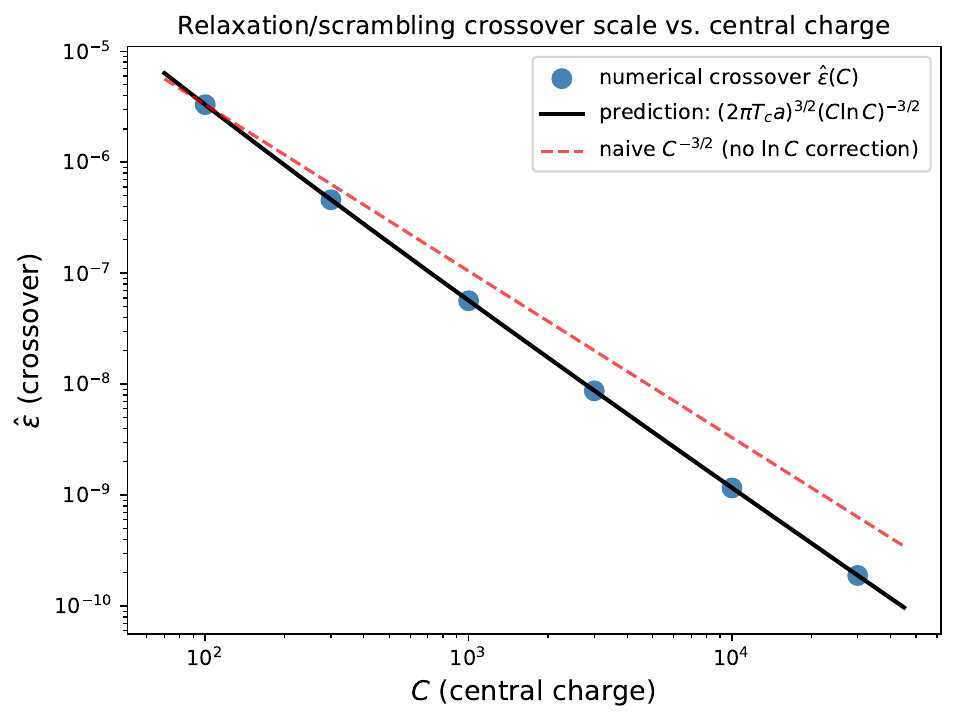}
\caption{Crossover reduced temperature $\hat{\epsilon}$ versus $C$. Points: numerical solution with the nonlinear free energy. Solid line: Eq.~\ref{eq:epsilon_hat}. Dashed line: a pure $C^{-3/2}$ law anchored at $C=10^2$}
\label{f11}
\end{figure}

The crossover point is then obtained by numerically solving Eq.~\ref{eq:crossover_condition} using a root-finding procedure in \(\epsilon\). In particular, the numerical calculation does not employ the critical-radius approximation \(x_e\simeq x_c\) used in deriving Eq.~\ref{eq:epsilon_hat}.

The resulting crossover temperatures are displayed in Fig.~\ref{f11}. The numerical data agree closely with the analytical expression in Eq.~\ref{eq:epsilon_hat}. The logarithmic correction in Eq.~\ref{eq:epsilon_asymptotic} is also important for describing the numerical behaviour. Over the range of central charges considered, a fit to a pure power law would yield an effective exponent different from \(-3/2\). Such a deviation does not indicate a failure of the analytical scaling relation; rather, it reflects the additional \((\ln C)^{-3/2}\) dependence. This distinction is illustrated by comparing the full analytical prediction with a pure \(C^{-3/2}\) scaling in Fig.~\ref{f11}.

\begin{figure}[h!]
\centering
\includegraphics[width=0.7\textwidth]{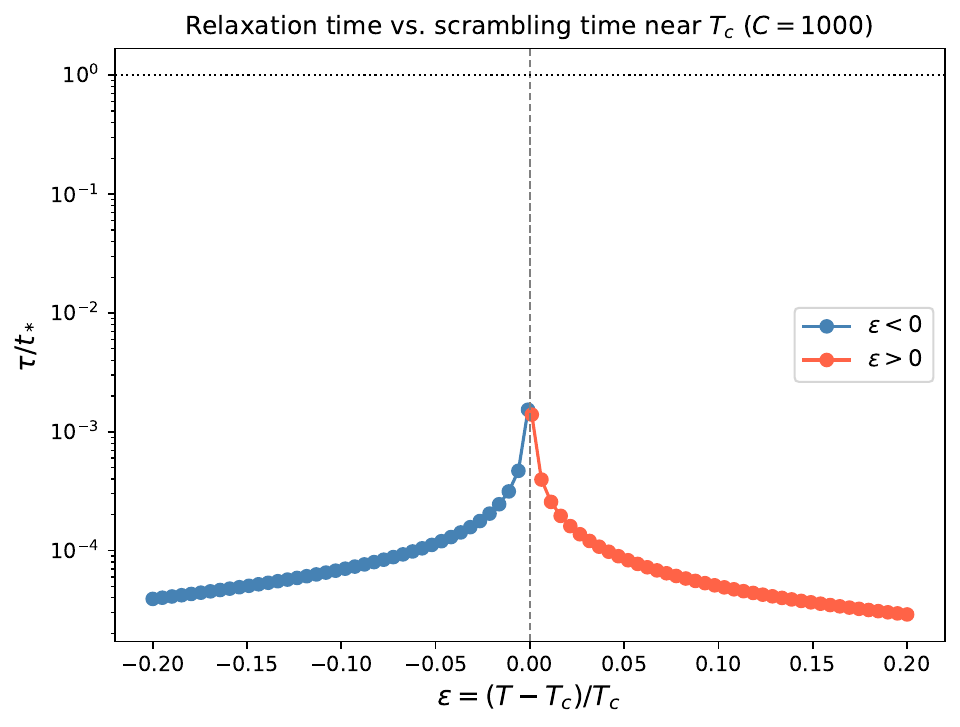}
\caption{Ratio $\tau/t_*$ as a function of $\epsilon$ at $C=10^3$ along the critical locus. The dotted line marks $\tau/t_*=1$.}
\label{f12}
\end{figure}

The hierarchy between the two timescales is further illustrated in Fig.~\ref{f12}, where we plot the ratio \(\tau/t_*\) as a function of the reduced temperature for a representative large value of the central charge. Away from criticality, this ratio remains much smaller than unity. As the critical point is approached, however, the relaxation time increases according to the power law \(|\epsilon|^{-2/3}\), while the scrambling time varies only logarithmically through the entropy and weakly through the temperature. The ratio therefore increases and eventually reaches unity at \(\epsilon=\hat\epsilon\).

These results suggest a hierarchy of dynamical timescales in the dual CFT. For \(\epsilon\gg\hat\epsilon\), the relaxation of the thermodynamic order parameter occurs on a timescale shorter than the scrambling time. In contrast, within the narrow region \(\epsilon\lesssim\hat\epsilon\), critical slowing down becomes sufficiently pronounced that the relaxation time exceeds the scrambling time. The width of this region decreases as \((C\ln C)^{-3/2}\), indicating that the regime in which the thermodynamic relaxation is slower than scrambling becomes increasingly restricted in the large-central-charge limit.

There is a further implication for the stochastic description employed in this work. The free-energy landscape approach treats the order parameter as a slow degree of freedom coupled to faster microscopic degrees of freedom that provide dissipation and fluctuations. The crossover scale identified above therefore offers a useful criterion for assessing the separation between these timescales. In particular, the regime \(\tau\gtrsim t_*\) identifies where the relaxation of the order parameter is no longer faster than the characteristic scrambling time of the microscopic theory. This comparison provides a motivation for examining the validity of the slow-variable description more carefully near criticality. We stress, however, that the present analysis does not derive the Langevin dynamics from a microscopic holographic calculation; the crossover should consequently be regarded as a timescale criterion rather than a proof of the validity or breakdown of the stochastic description.

Finally, the scaling in Eq.~\ref{eq:epsilon_asymptotic} is obtained with the dissipation coefficient held fixed. Since the amplitude \(a\) is proportional to \(\zeta\), a possible central-charge dependence of the dissipation coefficient would modify the crossover scaling. In particular, if \(\zeta\propto C^\kappa\), the leading dependence becomes

\begin{equation}
\hat\epsilon\sim
C^{-\frac{3}{2}(1-\kappa)}
(\ln C)^{-3/2}
\end{equation}

Thus, determining the central-charge dependence of \(\zeta\), for example through a calculation involving the lowest quasinormal mode of the dual gravitational system, is an important step towards establishing the microscopic dynamical origin of the crossover. We return to this issue in the concluding discussion.

\section{Conclusion and Discussion}\label{section6}

In this work, we investigate the phenomenon of critical slowing down associated with black hole phase transitions from the perspective of the dual CFT. Employing the holographic dictionary, we express the relevant bulk thermodynamic quantities in terms of their boundary counterparts for a charged AdS black hole. We identify the reduced horizon radius as the dynamical order parameter and examine its stochastic evolution in the vicinity of the critical point using the Langevin and Fokker-Planck formalisms. Our results reveal a pronounced slowing down of the relaxation dynamics in the dual CFT, manifested through a substantial enhancement of the autocorrelation time and the variance of the stochastic trajectories, together with a suppression of the smallest nonzero eigenvalue of the Fokker-Planck operator as criticality is approached. These observations are consistent with the critical slowing down previously reported in the bulk gravitational description \cite{slowing,Mozib}. The underlying mechanism can be attributed to the progressive flattening of the generalized free energy landscape near criticality, which weakens the effective restoring force and consequently leads to an enhancement of the characteristic relaxation timescale.

We further investigate the universality of the dynamical scaling behaviour in the dual CFT by analysing the autocorrelation time along different thermodynamic trajectories near the critical point. Our numerical results exhibit a power-law divergence of the form $\tau\sim|\epsilon|^{-2/3}$, with the extracted dynamical critical exponent in close agreement with the analytical prediction. Remarkably, this value is consistent with the scaling behaviour previously established for different black hole configurations in the bulk description \cite{Mozib}. This agreement suggests that the characteristic dynamical scaling associated with critical slowing down is also preserved in the holographically dual CFT description.

Going beyond the fixed-central-charge analysis, we have investigated the role of the central charge as an additional control parameter governing critical relaxation. By expanding the generalized free-energy landscape around the critical point, we identify a combined scaling field involving both the temperature deviation and the central-charge deviation. To leading order, the relaxation time depends on these two parameters through this single scaling variable, yielding a universal scaling form with the same critical exponent \(2/3\). Numerical calculations along multiple trajectories in the \((T,C)\) plane, performed using the full free-energy landscape, exhibit a clear collapse onto the analytical scaling prediction. This demonstrates that the leading critical relaxation behaviour is governed by a common scaling field even when temperature and central charge are varied simultaneously.

We have also examined the interplay between critical relaxation and quantum information scrambling in the dual CFT. By comparing the relaxation time with the scrambling time, we identify a crossover scale \(\hat{\epsilon}\) at which the two timescales become comparable. Along the critical locus, and for a central-charge-independent dissipation coefficient, our analysis yields the large-\(C\) behaviour
\begin{equation*}
\hat{\epsilon}\sim (C\ln C)^{-3/2}
\end{equation*}
The resulting crossover moves progressively closer to the critical point as the central charge increases, indicating that the regime in which thermodynamic relaxation becomes slower than scrambling becomes increasingly narrow in the large-\(C\) limit. The analytical prediction is further supported by numerical solutions obtained from the full nonlinear free-energy landscape, with the agreement improving systematically as \(C\) increases.

These findings provide a boundary perspective on the dynamical universality of black hole critical phenomena and reveal a nontrivial interplay between thermodynamic relaxation and microscopic scrambling in holographic CFTs. An important direction for future work is to determine the central-charge dependence of the effective dissipation coefficient from a microscopic holographic calculation, for instance through the analysis of quasinormal modes. Such an investigation would establish whether the crossover scaling obtained here persists beyond the fixed-dissipation approximation and would provide a more direct connection between the effective stochastic dynamics and the underlying microscopic theory.

\acknowledgments
We have used ChatGPT and Claude for numerical coding and draft-preparation workflow. All physics assumptions, numerical choices, and conclusions are the responsibility of the authors.
\appendix
\section{Numerical Implementation}\label{appendix A}

In this appendix, we describe the numerical framework adopted to investigate the stochastic dynamics of the reduced horizon radius $x$ and to determine the associated relaxation timescales. The analysis involves two complementary approaches: direct numerical integration of the Langevin equation and spectral analysis of the corresponding Fokker--Planck operator.

The stochastic evolution of the reduced horizon radius is governed by the overdamped Langevin equation
\begin{equation}
\zeta\frac{dx}{dt}
=
-\frac{\partial G(x)}{\partial x}
+\eta(t)
\label{eq:app_langevin}
\end{equation}
where $G(x)$ represents the generalized free energy landscape, $\zeta$ is the dissipation coefficient, and $\eta(t)$ denotes a Gaussian white-noise term. The statistical properties of the noise are specified by
\begin{equation}
\langle\eta(t)\rangle=0,
\qquad
\langle\eta(t)\eta(t')\rangle
=2\zeta T\delta(t-t')
\label{eq:app_noise}
\end{equation}
where $T$ is the ensemble temperature.

To generate the stochastic trajectories, we numerically integrate Eq.~\ref{eq:app_langevin} using the Heun predictor-corrector algorithm. Introducing the drift function
\begin{equation}
f(x)=-\frac{1}{\zeta}
\frac{\partial G}{\partial x}
\end{equation}
the discretized evolution over a time interval $h$ is implemented through a predictor step,
\begin{equation}
\bar{x}_{n+1}
=
x_n+f(x_n)h
+\sqrt{\frac{2Th}{\zeta}}\,\xi_n
\end{equation}
followed by the corrected update
\begin{equation}
x_{n+1}
=
x_n+\frac{h}{2}
\left[f(x_n)+f(\bar{x}_{n+1})\right]
+\sqrt{\frac{2Th}{\zeta}}\,\xi_n
\label{eq:app_heun}
\end{equation}
Here, $\xi_n$ represents independent random numbers sampled from a standard normal distribution. The same noise realization is used in the predictor and corrector steps. This procedure allows us to construct stochastic trajectories of the reduced horizon radius in the vicinity of equilibrium configurations. The resulting trajectories are subsequently used to evaluate the autocorrelation function and determine the corresponding relaxation time.

We also investigate the relaxation spectrum through the associated Fokker--Planck equation. The probability distribution $P(x,t)$ describing the stochastic evolution of the order parameter satisfies
\begin{equation}
\frac{\partial P(x,t)}{\partial t}=D\frac{\partial}{\partial x}\left[e^{-\beta G(x)}\frac{\partial}{\partial x}\left(e^{\beta G(x)}P(x,t)\right)
\right]
\label{eq:app_fp}
\end{equation}
where $D=T/\zeta$ is the diffusion coefficient and $\beta=1/T$. The equilibrium probability distribution is given by the Boltzmann form
\begin{equation}
P_{\mathrm{eq}}(x)
=
\frac{1}{Z}e^{-\beta\mathcal{G}(x)}
\label{eq:app_equilibrium}
\end{equation}
with $Z$ denoting the normalization factor.

To determine the relaxation spectrum, we employ a spectral decomposition of the probability distribution,
\begin{equation}
P(x,t)
=
e^{-\beta G(x)/2}
\psi(x)e^{-\lambda t}
\label{eq:app_spectral}
\end{equation}
Substitution into Eq.~\ref{eq:app_fp} transforms the Fokker-Planck equation into a Schr\"odinger-type eigenvalue problem,
\begin{equation}
-D\frac{d^2\psi}{dx^2}
+V_{\mathrm{eff}}(x)\psi
=
\lambda\psi
\label{eq:app_schrodinger}
\end{equation}
where the effective potential is determined by the generalized free energy landscape:
\begin{equation}
V_{\mathrm{eff}}(x)
=
D\left[
\frac{\beta^2}{4}
\left(\frac{\partial G}{\partial x}\right)^2
-\frac{\beta}{2}
\frac{\partial^2G}{\partial x^2}
\right]
\label{eq:app_veff}
\end{equation}
The eigenvalues $\lambda_n$ characterize the relaxation spectrum of the stochastic system. In particular, the vanishing eigenvalue corresponds to the stationary equilibrium distribution, whereas the smallest nonzero eigenvalue, $\lambda_1$, determines the slowest relaxation rate. Accordingly, the associated characteristic relaxation time is given by
\begin{equation}
\tau_{\mathrm{FP}}=\frac{1}{\lambda_1}
\label{eq:app_fp_time}
\end{equation}

The numerical determination of the Fokker-Planck eigenvalues is carried out following the spectral method described in Ref.~\cite{slowing}. Together, the Langevin simulations and the Fokker-Planck spectral analysis provide complementary characterizations of the relaxation dynamics and allow us to examine the emergence of critical slowing down in the dual CFT description.

\end{document}